\documentclass[journal]{IEEEtran}
\usepackage{url}
\usepackage{algorithm}
\usepackage{algorithmic}
\usepackage{ upgreek }
\ifCLASSINFOpdf
\usepackage[pdftex]{graphicx}
\else
\usepackage[dvips]{graphicx}
\fi
\usepackage{amsmath}
\usepackage{amsfonts} 
\usepackage{amssymb}
\usepackage{mathrsfs}

\usepackage{color}
\usepackage{ dsfont }
\usepackage[misc]{ifsym}
\usepackage{algorithmic}
\usepackage{array}
\usepackage{verbatim}
\ifCLASSOPTIONcompsoc
\usepackage[caption=false,font=normalsize,labelfont=sf,textfont=sf]{subfig}
\else
\usepackage[caption=false,font=footnotesize]{subfig}
\fi
\usepackage{amsmath}

\begin{document}

\title{Robust Data-driven Profile-based Pricing Schemes}

\author{Jingshi Cui, Haoxiang Wang, Chenye Wu, \emph{and} Yang Yu
\thanks{The authors are with the Institute for Interdisciplinary Information Sciences (IIIS), Tsinghua University, Beijing, China, 100084. C. Wu is the correspondence author. Email: chenyewu@tsinghua.edu.cn.}
}

\maketitle

\begin{abstract}
To enable an efficient electricity market, a good pricing scheme is of vital importance. Among many practical schemes, customized pricing is commonly believed to be able to best exploit the flexibility in the demand side. However, due to the large volume of consumers in the electricity sector, such task is simply too overwhelming. In this paper, we first compare two data driven schemes: one based on load profile and the other based on user's marginal system cost. Vulnerability analysis shows that the former approach may lead to loopholes in the electricity market while the latter one is able to guarantee the robustness, which yields our robust data-driven pricing scheme. Although $k$-means clustering is in general NP-hard, surprisingly, by exploiting the structure of our problem, we design an efficient yet optimal $k$-means clustering algorithm to implement our proposed scheme.
\end{abstract}


\IEEEpeerreviewmaketitle

\section{Introduction}\label{Introduction}

The legacy power grid has worked admirably for over a century under the notion that electricity access is a human right \cite{worldbankReport}. This notion makes the power system operators often overlook the opportunities brought by \textit{extensively} exploiting the value of customers. The challenge also comes from the large volume of customers in the system. It is simply too overwhelming for the operator to design customized pricing schemes for all end users, especially with time varying energy consumption patterns. At its core, this is a big data issue.

Powered by the wide deployment of smart meters, the advanced data analytic methods and machine learning algorithms ease the process of customized pricing by providing promising solutions of using big data to inform better price decisions to the system operator.

\begin{figure}[!t]
\centering
\includegraphics[width=2.5 in]{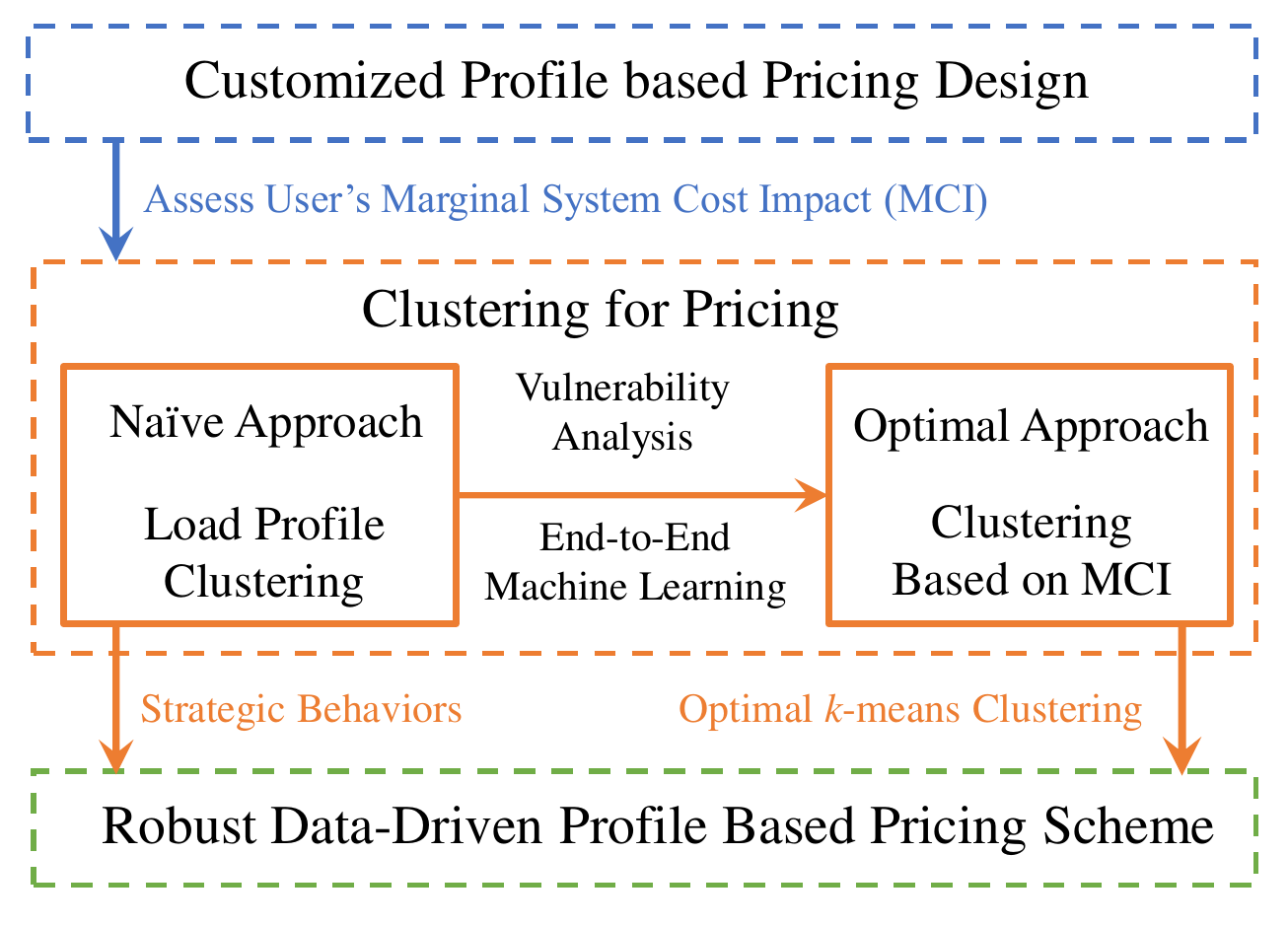}
\caption{Structure in Designing the Robust Pricing Scheme}
\label{framework}
\end{figure}

To exemplify this solution for the electricity sector, in this paper, we study a microgrid scenario, where the microgrid operator sets the real time price according to the marginal generation cost. Inspired by Yu $\textit{et al.}$ in \cite{Yang2017Good}, we define a metric $MCI$, based on the individual load profiles, to reflect each user's value in the microgrid operation. This metric allows us to employ $k$-means clustering algorithm to design the customized pricing scheme. Unlike most previous studies, which conduct the $k$-means clustering based on user's load profile, we adopt the notion of end-to-end machine learning, which yields a robust data-driven pricing scheme. Figure \ref{framework} visualizes the structure of this work.

\subsection{Literature Review}

To the best of our knowledge, we are the first to design a robust data-driven profile-based pricing scheme for the electricity sector. We identify three groups of closely related literature.

The first related research direction investigates using machine learning techniques for better power system operation, mostly with a focus on load prediction. Just to name a few, Kong $\textit{et al.}$ exploit the temporal characteristics in residential load and propose an LSTM (Long Short-Term Memory) load forecasting method in \cite{Kong2017Short}. Many deep learning models have also been employed for more accurate load prediction, including pooling-based deep recurrent neural network in \cite{Shi2018Deep}, FFNN (Feed-Forward Neural Network) in \cite{Wei2019Prediction}, DBN (Deep Belief Network) in \cite{Dedinec2016Deep}, etc. Chen $\textit{et al.}$ seek to integrate domain knowledge in different neural network building blocks in \cite{Chen2018Short}. Researchers have also used machine learning to solve scheduling or dispatch problems for electricity market. For example, Mocanu $\textit{et al.}$ propose a deep reinforcement learning method to conceive an online optimization for building energy management systems in \cite{Mocanu2017On}. 

Another line of relevant research exploits opportunities in designing better pricing schemes for demand side. Based on utility maximization, Samadi $\textit{et al.}$ propose an optimal real-time pricing algorithm for demand side management in \cite{Samadi2010Optimal}. By combining hybrid particle swarm optimizer with mutation algorithm, Xu $\textit{et al.}$ design a data-driven pricing scheme to help the utility minimize peak demand in \cite{Xu2017Data}. Exploiting price elasticity of electricity consumption, Yu $\textit{et al.}$ develop a parametric time-utility model to obtain the optimal real time price and maximize the social welfare in \cite{Yu2012A}. Qian $\textit{et al.}$ develop an SAPC (Simulated-Annealing-based Price Control) algorithm to reduce the peak-to-average load ratio via a two-stage optimization in \cite{Li2013Demand}.

Lastly, the robustness of clustering algorithms has also been well investigated. For example, Georgogiannis $\textit{et al.}$ use outlier detection to conduct the robustness analysis for quadratic $k$-means in \cite{georgogiannis2016robust}. Alok $\textit{et al.}$ overview the cluster validity measures and analyze the robustness of such indices in \cite{Validity}. Yan $\textit{et al.}$ investigate the consistency and robustness of two kernel-based clustering algorithms: semidefinite programming relaxation for kernel clustering and kernel singular value decomposition in \cite{yan2016robustness}. Xu $\textit{et al.}$ design a robust and sparse fuzzy $k$-means clustering algorithm by incorporating a robust function to handle outliers in \cite{xu2016robust}. Li $\textit{et al.}$ introduce the concept of robust cluster objective function to enhance the robustness of clustering algorithms in \cite{li2016robust}.

\subsection{Our Contributions}
In pursuit of a robust data driven pricing scheme, the principal contributions of our work can be summarized as follows:

\begin{itemize}
    \item \textit{End-to-End Machine Learning}: We adopt the notion of end-to-end machine learning to design the optimal $k$-means clustering algorithm, which highlights that the vulnerability of many data-driven pricing schemes (e.g., the one in our recent work \cite{cui2019vulnerability}) is due to the clustering criteria selection.
    \item \textit{Characterize Robustness}: Following the definitions of strategic behaviors in \cite{cui2019vulnerability}, we propose a smoothness requirement to characterize robustness. This serves as the guideline to design robust data-driven pricing scheme. Surprisingly, although smoothness is a \textit{global} concept, we prove that a $\textit{local}$ criteria for each cluster implies this $\textit{global}$ concept.
    \item \textit{Optimal $k$-means Clustering}: Although $k$-means clustering in general is NP-hard \cite{dasgupta2008hardness}, we exploit the structure of our problem and design a greedy algorithm to conduct $k$-means clustering with the smoothness requirement. We further prove that to achieve the smoothness requirement, the greedy algorithm is optimal in terms of minimizing the number of clusters.
\end{itemize}

The rest of the paper is organized as follows. We introduce our system model and identify each user's marginal system cost impact ($MCI$) in Section \ref{SystemModel}. Then, Section \ref{Vulnerability} examines the economic loophole created by common $k$-means clustering. Inspired by such vulnerability analysis, in Section \ref{Robustness}, we define robustness and propose that adopting the notion of end-to-end machine learning leads to a robust scheme: clustering based on $MCI$. We further design the greedy $k$-means clustering algorithm and prove its optimality in Section \ref{RobustPricing}. Simulation studies are conducted in Section \ref{Simulaiton} for further insights into the proposed scheme. Concluding remarks and future directions are given in Section \ref{Conclusion}.

\section{System Model}\label{SystemModel}

\begin{figure}[!t]
\centering
\includegraphics[width=2.5 in]{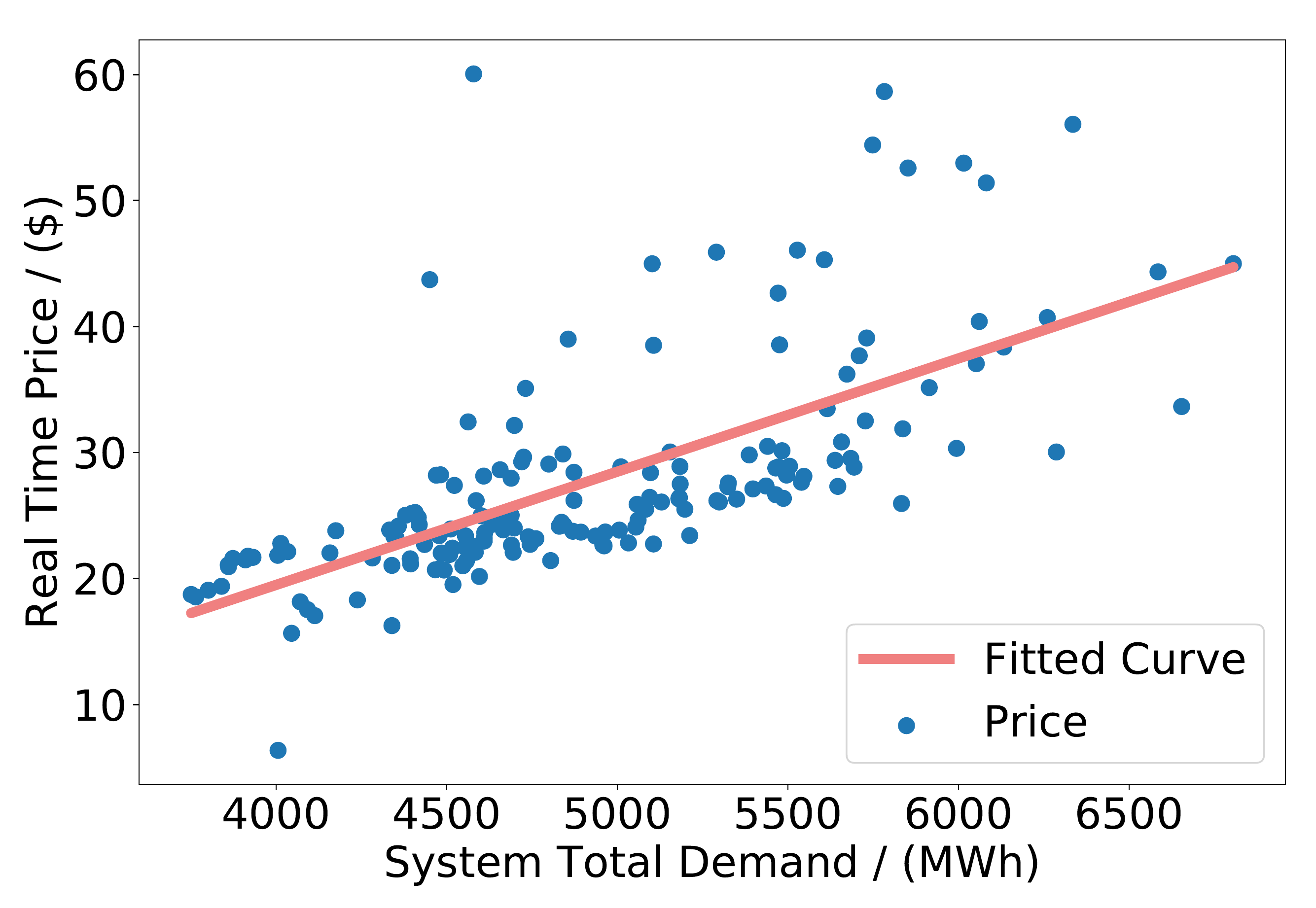}
\caption{Relationship Between Price and Demand}
\label{PriceDemand}
\end{figure}

We consider a microgrid of $N$ users where the microgrid energy acquisition is maintained by an operator. Denote the total energy consumption in the microgrid at time $t$ by $L_{t}$. Assume the cost to acquire energy is of a quadratic form, i.e.,
\begin{equation}\label{CostFunction}
\begin{aligned}
C(L_{t})=&\frac{1}{2}a\cdot L_{t}^{2}+b\cdot L_{t}+c,\,\,\,\forall 1\leq t\leq T,\\
s.t. \,\,\,&L_{t}=\sum\nolimits_{i=1}^{N}l_{i}^{t},
\end{aligned}
\end{equation}
where $l_{i}^{t}$ denotes user $i$'s energy consumption at time $t$. In such a model, the microgrid operator could simply set the electricity price $p(t)$ at time $t$ according to the marginal cost:
\begin{equation}\label{price}
p(t)=\frac{\partial{C(L_{t})} }{\partial{L_{t}}}=aL_{t}+b,\,\,\,\forall 1\leq t\leq T.
\end{equation}

\noindent $\textbf{Remark}$: When $N$ is large, the energy consumption in the microgrid forms a perfect competition (i.e. each user is a price taker) and the price determined by Eq. (\ref{price}) is efficient. We adopt such a model in that it also allows us to understand the case when the price taker assumption does not hold, though out of the scope of this paper. Furthermore, the quadratic form of cost functions (i.e. linear form of prices) is a common assumption \cite{martinez2003adapting}, which can be verified by real data. Figure \ref{PriceDemand} suggests that the real time price ($p(t)$) and the system total demand exhibit a linear relationship, using the PJM data \cite{DemandPrice}.

\subsection{Marginal System Cost Impact}
We follow the metric, marginal system cost impact (denoted by $MCI$), first proposed by Yu $\textit{et al.}$ \cite{Yang2017Good} to assess each individual user's system impact.

In the daily operation of the microgrid, the total cost for the microgrid operator can be obtained as follows:
\begin{equation}
    C_{T}(\mathbf{L})=\sum\nolimits_{t=1}^{T}C(L_{t}),
\end{equation}
where $\mathbf{L}=(L_{1},...,L_{T})$. If the operator implements an hourly price, then $T$ is simply $24$ for the daily operation.

The $MCI$ is characterized by each user's own load profile. For each user $i$, denote its daily load profile by $\mathbf{l}_{i}=(l_{i}^{1},...,l_{i}^{T})$. Then, we have
\begin{equation}
\begin{aligned}
MCI_{i}&=\lim_{\Delta \rightarrow 0}\frac{C_{T}\left(\mathbf{L}+\frac{\Delta \mathbf{l}_{i}}{||\mathbf{l}_{i}||_{1}}\right)-C_{T}(\mathbf{L})}{\Delta }\\
&=\lim_{\Delta \rightarrow 0}\frac{\sum_{t=1}^{T}(aL_{t}+b)\frac{\Delta l_{i}^{t}}{||\mathbf{l}_{i}||_{1}}+\left (  \frac{\Delta l_{i}^{t}}{||\mathbf{l}_{i}||_{1}}\right )^{2}}{\Delta }\\
&=\sum\nolimits_{t=1}^{T}(aL_{t}+b)\frac{l_{i}^{t}}{||\mathbf{l}_{i}||_{1}}\\
&=\sum\nolimits_{t=1}^{T}p(t)\frac{l_{i}^{t}}{||\mathbf{l}_{i}||_{1}}.
\end{aligned}
\end{equation}

Note that, by adopting $MCI_{i}$ as the customized daily electricity rate for user $i$, its daily electricity cost is as follows:
\begin{equation}
\begin{aligned}
MCI_{i}\cdot ||\mathbf{l}_{t}||_{1}&=\left (  \sum_{t=1}^{T}p(t)\frac{l_{i}^{t}}{||\mathbf{l}_{t}||_{1}}) \right )\cdot ||\mathbf{l}_{t}||_{1}\\
&=\sum\nolimits_{t=1}^{T}p(t)\cdot l_{i}^{t}.
\end{aligned}
\end{equation}

This is a remarkable result! It shows that $MCI_{i}$, as a customized uniform price, maintains the same system efficiency as the prices determined by the marginal cost (i.e., $p(t)$'s). Moreover, $MCI_{i}$ reflects the true value/cost of user $i$ to the whole microgrid system. 

However, we want to emphasize that our subsequent analysis does not rely on the quadratic cost functions. All of our conclusions hold as long as the first and second order derivatives of $C(L_{t})$ exist, and its second order derivative is bounded.

This result again verifies the conclusion in \cite{Yang2017Good} that each individual user's load profile should uniquely determine the user's electricity rate, which serves as the basis for data-driven pricing design.

\subsection{Dataset Overview}\label{DataOverview}
Before diving into the details of discussion on the data-driven pricing, we first provide a brief overview on the datasets, which we will use throughout the paper.

We consider two types of demand profiles, residential users and commercial building users: one with more diversity and the other with an inherent pattern.

Empirical studies have shown that the load profiles across populations show similar patterns \cite{rasanen2010data}. Hence, to enlarge the two datasets for more informative clustering results, we combine the data in different days. More precisely,

\begin{enumerate}
    \item Residential Dataset: We use the Pecan Street dataset \cite{DemandData} with more than $300$ residential load profiles from $2016.1.1$ to $2016.2.10$. We combine all the load profiles (excluding some profiles with missing data) into a single residential dataset of $7,699$ profiles.
    \item Commercial Building Dataset: We use the Energy Plus dataset \cite{BuildingData} with $4,820$ commercial building load profiles across US from $2007.1.2$ to $2007.1.12$. By combining all the data (again, excluding some profiles with missing data) to form a large dataset, we obtain our commercial building dataset of $48,200$ profiles.
\end{enumerate}

As suggested by the definition of $MCI$, besides calculating the total load $L_{t}$, we normalize the profiles in two datasets to highlight the observation that user's load profile uniquely determines its electricity rate. Mathematically, for each user $i$'s daily energy consumption $\mathbf{l}_{i}$, we define its normalized consumption $\mathbf{d}_{i}$ according to the $l_{1}-$norm:
\begin{equation}
\mathbf{d}_{i}=\frac{\mathbf{l}_{i}}{||\mathbf{l}_{i}||_{1}}=\frac{\mathbf{l}_{i}}{\sum_{t=1}^{T}l_{i}^{t}}.
\end{equation}

\section{Vulnerability of Common Clustering}\label{Vulnerability}

One straightforward way for clustering is to conduct the $k$-means clustering based on each individual's load profile. However, since the users in the same cluster do not share exactly the same load profile, this data-driven pricing approach may create loopholes allowing certain users to bypass to other clusters for a better retail price.

In this section, we revisit the definition of disguising in \cite{cui2019vulnerability}, and provide more empirical evidence to support that strategic behaviors do exist under this pricing scheme.

\subsection{Definition of Disguising}

Users may strategically change their load profiles for a better price, especially for those who lie on the boundary of two clusters. They can conduct minimal load profile modifications to switch to another cluster, and potentially for a better price. We define such behaviors as disguising.

We denote the center (load profile) of cluster $j$ by $\mathbf{c}_{j}$, and denote the cluster that user $i$ belongs to by a function $u(i)$. To understand the strategic behavior, we need to examine the effort that user $i$ needs to spend on moving to another cluster $n\neq u(i)$. This effort can be characterized by a scalar $\mu_{i,n}$. If user $i$ can successfully switch to cluster $n$, the following condition holds:
\begin{equation}\label{SwitchCondition}
    ||(1-\mu_{i,n})\mathbf{d}_{i}+\mu_{i,n}\mathbf{c}_{n}-\mathbf{c}_{u(i)}||_{1}\geq ||(1-\mu_{i,n})(\mathbf{d}_{i}-\mathbf{c}_{n})||_{1}.
\end{equation}
Hence, the minimal effort that user $i$ needs to spend for disguising can be solved by an optimization problem:
\begin{equation}\label{MinMove}
\begin{aligned}
\min_{n\neq u(i)}\,\,& \inf \mu_{i,n} \\
s.t.\,\,\,\,&||(1-\mu_{i,n})\mathbf{d}_{i}+\mu_{i,n}\mathbf{c}_{n}-\mathbf{c}_{u(i)}||_{1}\\
&\ \ \ \ \geq ||(1-\mu_{i,n})(\mathbf{d}_{i}-\mathbf{c}_{n})||_{1},\\
&p_{n}< p_{u(i)},
\end{aligned}
\end{equation}
where $p_{n}$ represents the price of cluster $n$, i.e., the $MCI$ for cluster $n$'s center profile $\mathbf{c}_{n}$.

Denote $CR$ to be the minimal objective value to problem ($\ref{MinMove}$). We can formally characterize disguising as follows.

\vspace{0.1cm}
\noindent $\textbf{Definition 1}$: User $i$ has the ability to disguise if
\begin{equation}\label{theta}
    CR_{i}\leq \theta.
\end{equation}

\noindent $\textbf{Remark}$: In this \textit{parametric} definition, the user's strategic behavior can be measured through a threshold $\theta$. We also want to emphasize that by considering the load profile modification based on the normalized profile, we intentionally ignore user's possibility of disguising via reducing energy consumption, which allows us to focus on understanding the potential of shifting load profile for disguising.

\subsection{Empirical Evidence: Clustering Results}

\begin{figure}[!t]
\centering
\subfloat[Residential Dataset]{\includegraphics[width=3in]{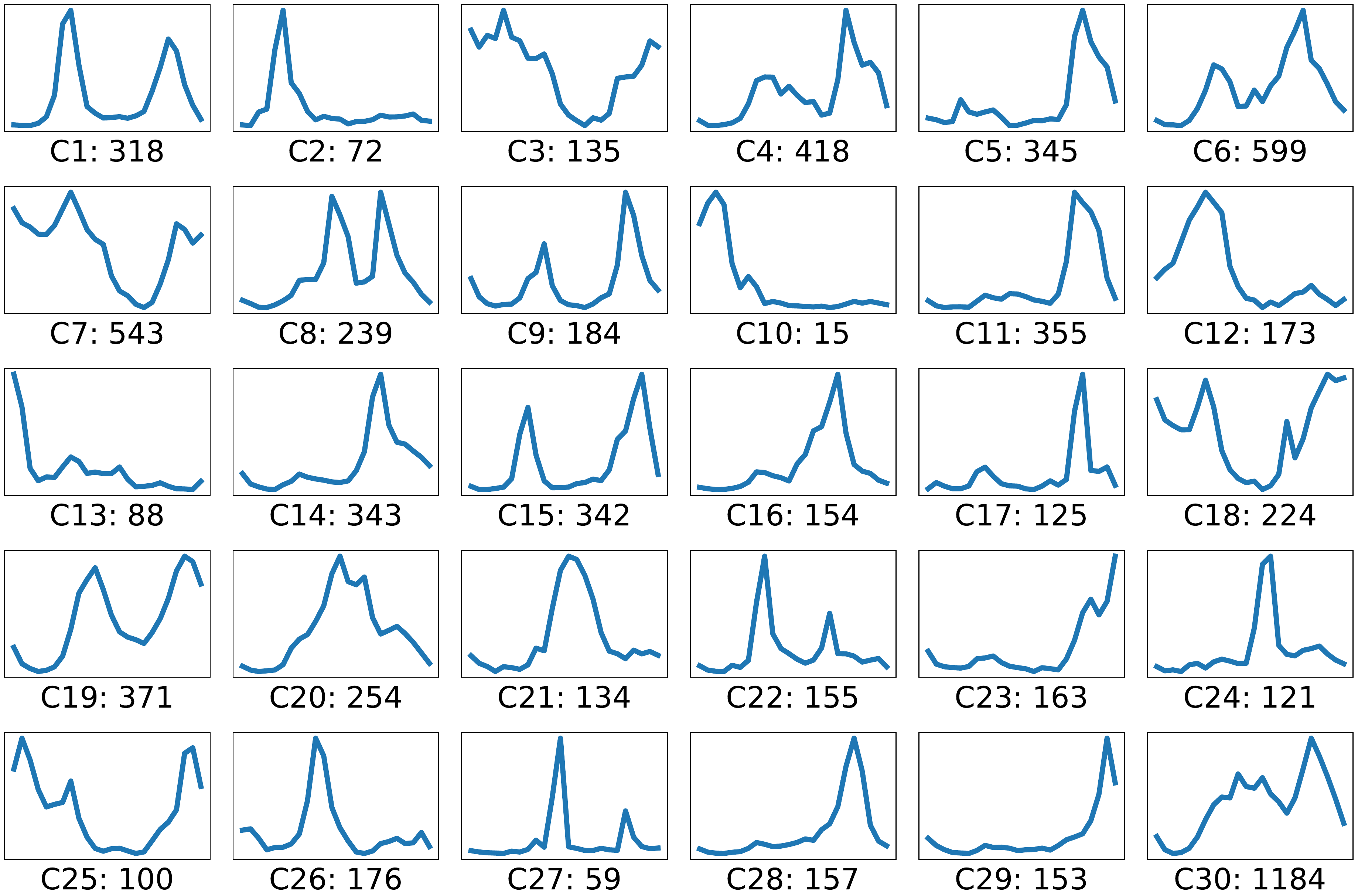}%
\label{UserCluster}}
\hfil
\subfloat[Commercial Building Dataset]{\includegraphics[width=3in]{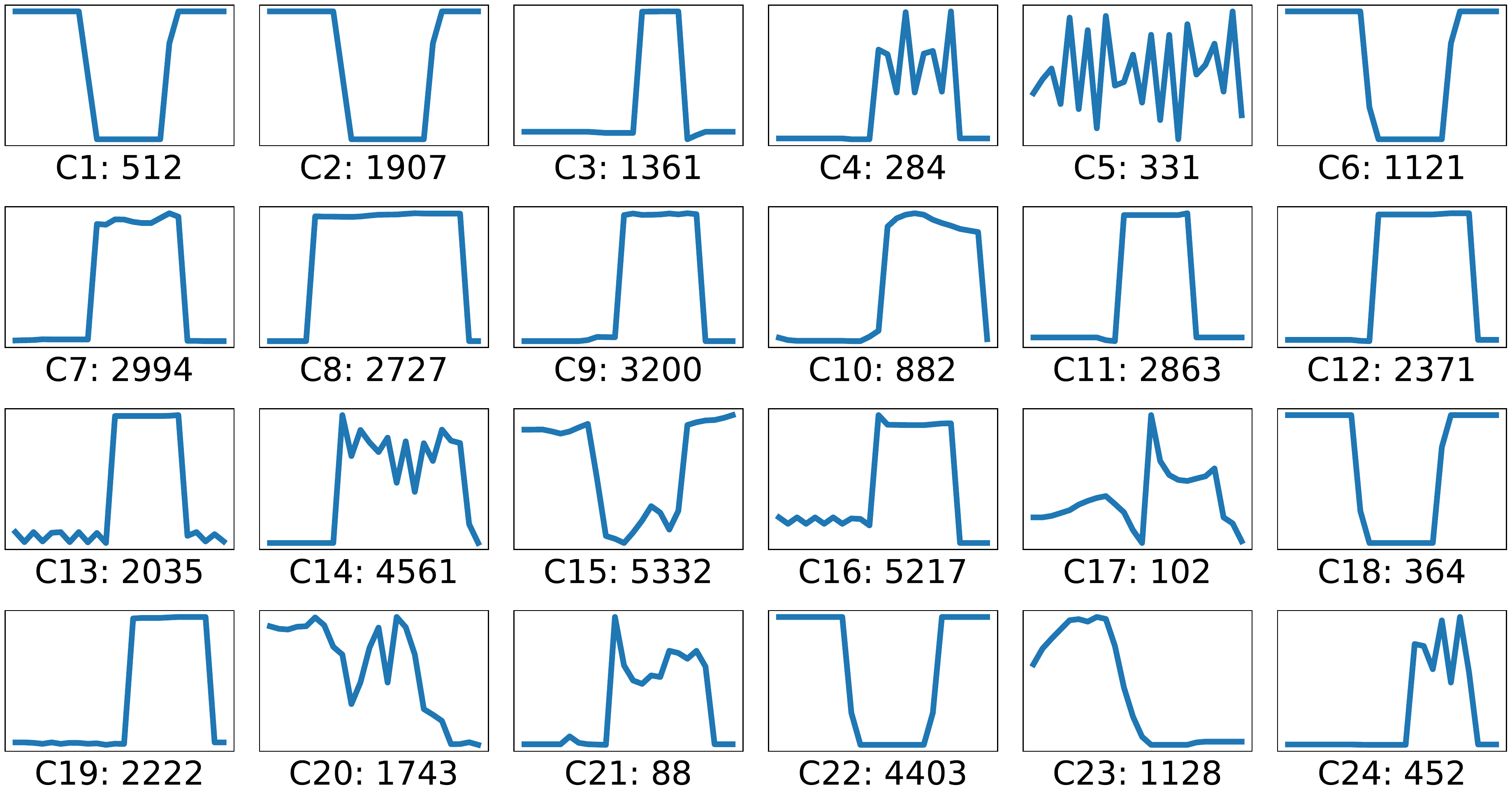}%
\label{BuildingCluster}}
\caption{Demand Profile of Each Cluster Center}
\label{ClusterResult}
\end{figure}

We first conduct the $k$-means clustering for the two datasets. We choose a small $k$ for better illustration of the subsequent analysis on strategic behaviors. Fig \ref{ClusterResult}\subref{UserCluster} and \ref{ClusterResult}\subref{BuildingCluster} visualize the clustering results for the two datasets, respectively.

The central profiles show interesting patterns in the two types of users. For example, cluster $8$ in Fig. \ref{ClusterResult}\subref{UserCluster} and cluster $4$ in Fig. \ref{ClusterResult}\subref{BuildingCluster} seem more like the classical load pattern with $2$ peaks: one in the noon time and the other at night. We can observe that there are many more types of users, including those who are more active at night (e.g. cluster $11$ in Fig. \ref{ClusterResult}\subref{UserCluster}, and cluster $22$ in Fig. \ref{ClusterResult}\subref{BuildingCluster}). 

Note that, although the size of the building dataset is almost six times that of the residential user dataset, it leads to fewer clusters. Even within the $24$ clusters, one may observe the inherent pattern of commercial buildings. This observation corresponds to our descriptions for the two datasets in Section \ref{DataOverview}. Residential load profiles are more heterogeneous while commercial building profiles exhibit more homogeneity.

\subsection{Empirical Evidence: Strategic Behaviors}

To quantify the existence of strategic behaviors, we can define the number of users, having the ability to disguise, in each cluster $n$ as follows \cite{cui2019vulnerability}:
\begin{equation}\label{N}
N_{n}(\theta):=\sum\nolimits_{i,u(i)=n}I(CR_{i}\leq \theta),
\end{equation}
where $I(\cdot )$ is the indicator function.

\begin{figure*}[!t]
\centering
\subfloat[Residential Users]{\includegraphics[width=3in]{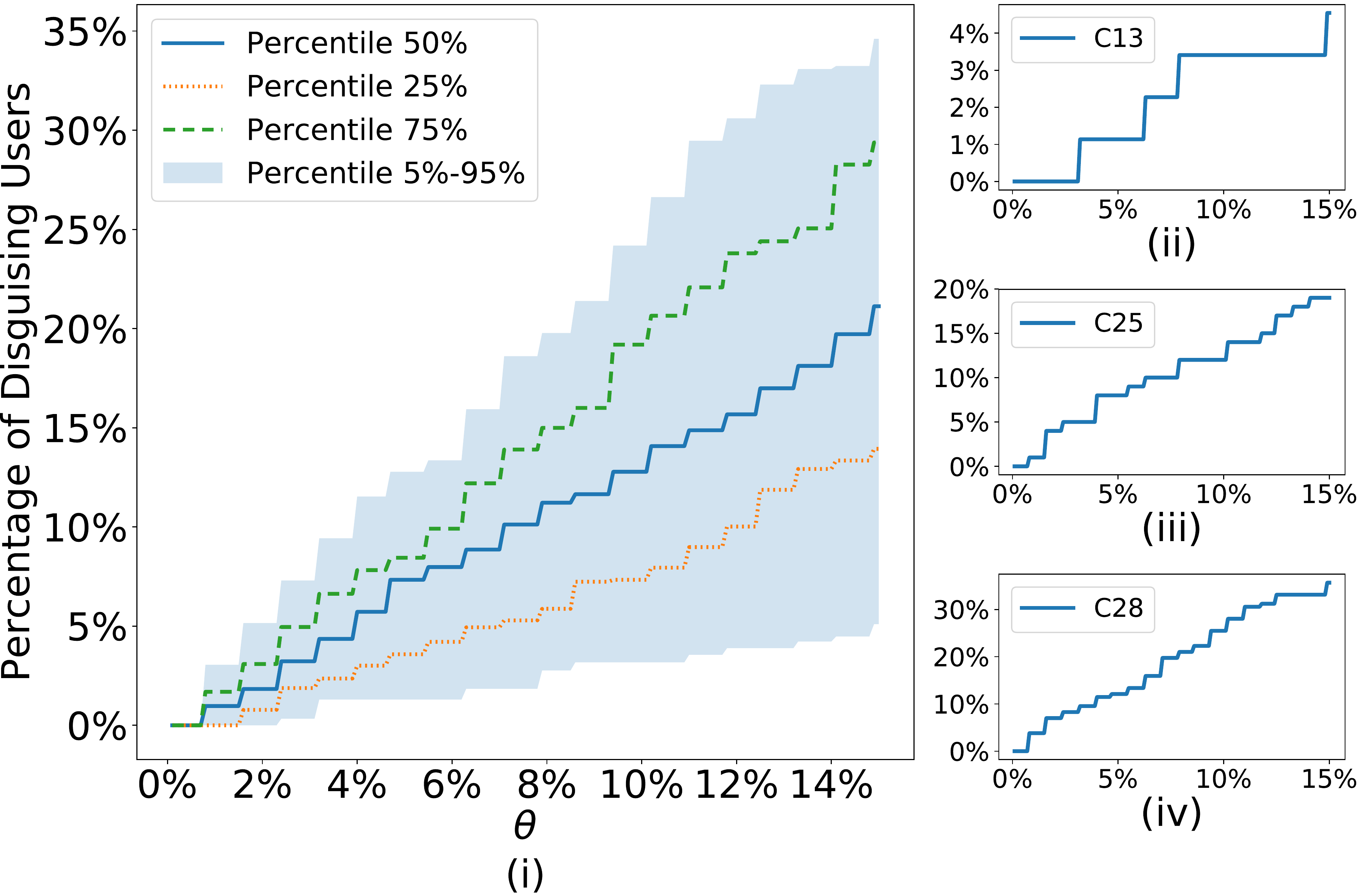}%
\label{UserNPercent}}
\hfil
\subfloat[Commercial Building Users]{\includegraphics[width=3in]{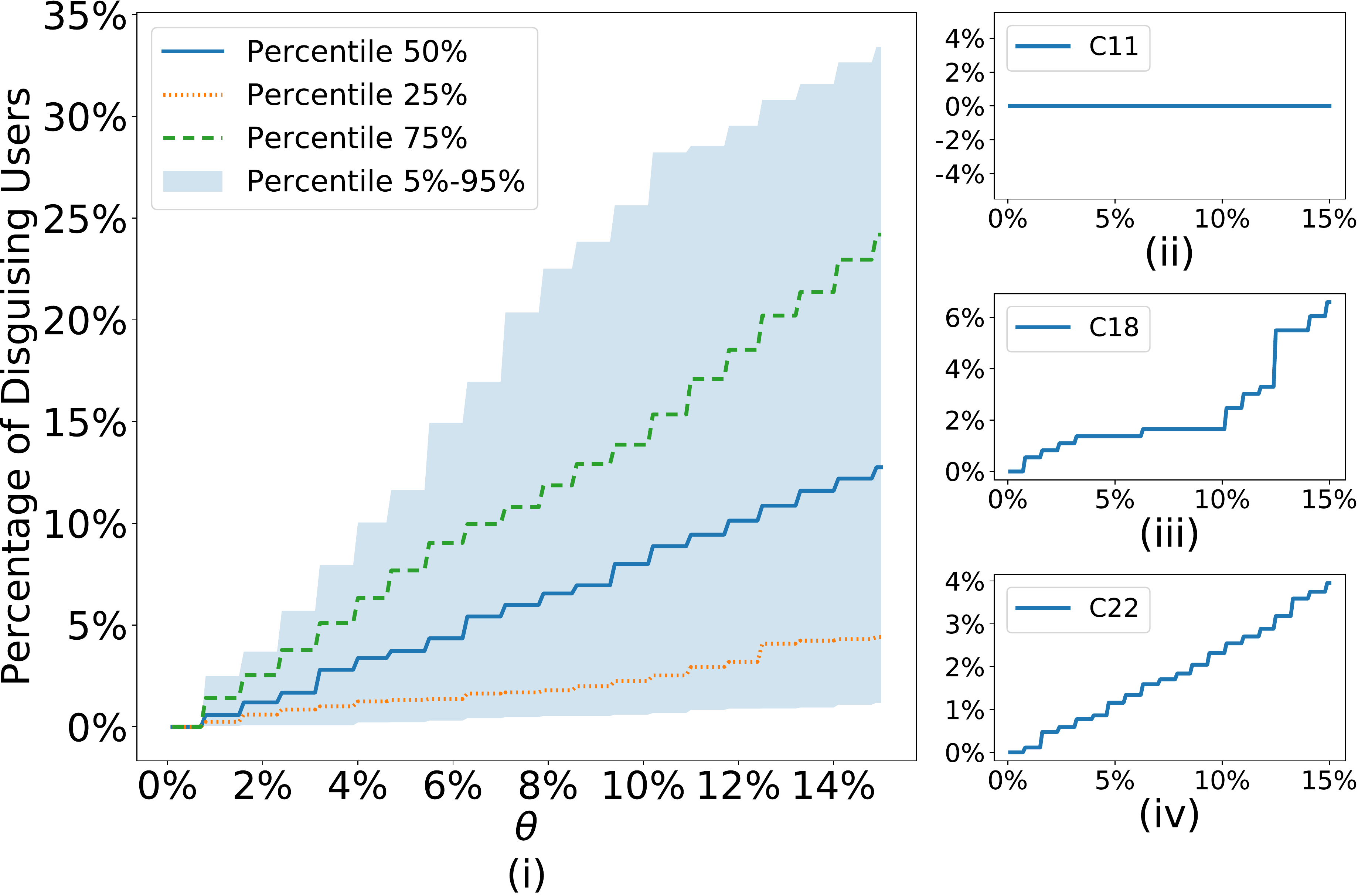}%
\label{BuildingNPercent}}
\caption{Percentage of Strategic Users Evolving with $\theta$}
\label{NPercent}
\end{figure*}

Figure \ref{NPercent} plots how the percentage of potential strategic users varies with $\theta$. Figure \ref{UserNPercent}(i) and \ref{BuildingNPercent}(i) demonstrate the statistical trends for the two datasets, which are quite intuitive. With a relaxed definition of disguising (larger $\theta$), more users can become strategic. Yet, it is interesting to note that, in both datasets, users can disguise only if they change at least $0.8\%$ of their load profiles.

We show more results in the subfigures of Fig. \ref{NPercent} to highlight the heterogeneity in clusters. There are rather stable clusters, e.g. cluster $11$ in the commercial building dataset (as shown in Fig. \ref{BuildingNPercent}(ii)). And there are also rather unstable ones, e.g. cluster $28$ in Fig. \ref{UserNPercent}(iv), and cluster $22$ in Fig. \ref{BuildingNPercent}(iv).

\section{Robust Data-Driven Pricing: Theory}\label{Robustness}

To understand why strategic behaviors exist in the straightforward $k$-means clustering, we evaluate user's potential benefits of disguising.

Suppose user $i$ has disguised itself to be a cluster $n$ user, with a disguised profile $\mathbf{\widetilde{d}}_{i,n}$, where
\begin{equation}
\mathbf{\widetilde{d}}_{i,n}=((1-\mu_{i,n})\mathbf{d}_{i}+\mu_{i,n}\mathbf{c}_{n}).
\end{equation}

Then its potential benefit can be measured by the difference in electricity bills $b_{i,n}$:
\begin{equation}
    b_{i,n}(\mu_{i,n})=p_{u(i)}\|\mathbf{d}_{i}\|_{1}-p_{n}\|\mathbf{\widetilde{d}}_{i,n}\|_{1}.
\end{equation}
Recall that $p_{n}$ denotes the electricity rate for users in cluster $n$. Note that, since the load profiles have been normalized, $b_{i,n}$ measures the unit benefit of disguising. Mathematically, since
$\|\mathbf{d}_{i}\|_{1}=\|\mathbf{\widetilde{d}}_{i,n}\|_{1}=1$, we have 
\begin{equation}
    b_{i,n}(\mu_{i,n})=p_{u(i)}-p_{n}.
\end{equation}

Hence, each user's maximal potential to disguise, denoted by $b_{i}$, can be defined as follows
\begin{equation}\label{maxbenefit}
    b_{i,n}(\mu_{i,n})=\max_{n\in \mathcal{K}_{i},n\neq n(i)}\{p_{u(i)}-p_{n}\},
\end{equation}
where $\mathcal{K}_{i}$ denotes the set of clusters that user $i$ can successfully disguise itself with $CR_{i}\leq\theta$.

\subsection{Definition of Smoothness}
The potential benefits of disguising in Eq. (\ref{maxbenefit}) motivate us to define smoothness to ensure a robust data-driven scheme.

\vspace{0.1cm}
\noindent $\textbf{Definition 2}$: The $k$-means clustering is $\delta-$smooth if for any user $i$ and its associated disguising set $\mathcal{K}_{i}$, the following condition holds:
\begin{equation}\label{deltasmooth}
    |p_{u(i)}-p_{n}| \leq \delta,\forall n\in \mathcal{K}_{i}.
\end{equation}

Obviously, $\delta-$smoothness guarantees the bound of robustness. One straightforward way to ensure such smoothness is to design a supervised load profile based $k$-means clustering with the smoothness requirement. 

\subsection{$k$-means Clustering with Smoothness Guarantee}
In this section, we want to show that, although smoothness is a global concept, it can be guaranteed by local properties. More precisely, if a $k$-means clustering could ensure that for each user $i$, its $MCI_{i}$ is close to $p_{u(i)}$, then, this $k$-means clustering is smooth. Formally, we can prove the following theorem:

\vspace{0.1cm}
\noindent $\textbf{Theorem 1}$: Suppose a $k$-means clustering algorithm can guarantee that
\begin{equation}\label{delta}
    |MCI_{i}-p_{u(i)}| \leq \rho, \forall i \in u(i),
\end{equation}
then the clustering is $\rho(1+\frac{1}{1-\theta})-$smooth.

\vspace{0.2cm}
\noindent $\textbf{Proof}$:
Suppose user $i$ successfully disguises itself as cluster $n$ user with a profile $\mathbf{\widetilde{d}}_{i,n}$. The condition in the theorem implies that,
\begin{equation}\label{pn}
\begin{aligned}
&\left|p_{n}-\sum\nolimits_{t=1}^{T}p(t)((1-\mu_{i,n})d_{i}^{t}+\mu_{i,n}c_{n}^{t})\right|\\
=&\left |\sum\nolimits_{t=1}^{T}p(t)(1-\mu_{i,n})(c_{n}^{t}-d_{i}^{t})\right|\\
=&|(1-\mu_{i,n})(p_{n}-MCI_{i})|\\
\leq &\rho .
\end{aligned}
\end{equation}

Hence, for any cluster $n\in \mathcal{K}_{i}$, we have
\begin{equation}
\begin{aligned}
|p_{n}-p_{u(i)}|\leq &|p_{n}-MCI_{i}|+|MCI_{i}-p_{u(i)}|\\
\leq & \frac{\rho}{1-\mu_{i,n}}+\rho\\
\leq &\rho \left ( 1+\frac{1}{1-\theta} \right ).
\end{aligned}
\end{equation}
The second inequality is due to Eq. (\ref{pn}) and the last inequality is due to the parametric definition of disguising. Thus, we complete the proof.
\hfill$\blacksquare$ 
\vspace{0.1cm}

The only remaining handle is to design an efficient $k$-means clustering algorithm with smoothness guarantee.

\section{Robust Data-Driven Pricing: Implementation}\label{RobustPricing}

In this section, we first introduce the most straightforward supervised $k$-means clustering implementation, and then propose our greedy algorithm. To our surprise, although the general $k$-means clustering is NP-hard, the proposed greedy algorithm is efficient and optimal.

\subsection{Supervised $k$-Means Clustering}

Given the outcome of profiled based $k$-means clustering (see the algorithm in \cite{cui2019vulnerability} for an instance), we can modify the outcome (details are provided in Algorithm $1$) to ensure the smoothness requirement. We term this algorithm as S$k$C for short.

\renewcommand{\algorithmicrequire}{\textbf{Input:}}  
\renewcommand{\algorithmicensure}{\textbf{Output:}} 

\begin{algorithm}[!ht]
\caption{Supervised $k$-means Clustering (S$k$C)}
\begin{algorithmic}[1]
\REQUIRE User data set $C=\left \{ i|i=1,...,n \right \}$, price data $p(t)$
\ENSURE Clusters $C_{1},...,C_{k}$
\STATE $m=\min_{i} MCI_{i}$
\STATE $M=\max_{i} MCI_{i}$
\IF {$M-m < 2 \rho$}
\RETURN $C$ as a single cluster
\ELSE
\STATE $C_{1}=\varnothing $, $C_{2}=\varnothing $
\FOR{$i=1$ to $n$}
\IF {$|MCI_{i}-m|\leq |MCI_{i}-M|$}
\STATE $C_{1}=C_{1} \cup \left \{ i \right \}$
\ELSE
\STATE $C_{2}=C_{2} \cup \left \{ i \right \}$
\ENDIF
\ENDFOR
\ENDIF
\IF{$|\max_{i\in C_{1}}MCI_{i}-\min_{i\in C_{1}}MCI_{i}|>2 \rho$}
\IF{$|\max_{i\in C_{2}}MCI_{i}-\min_{i\in C_{2}}MCI_{i}|>2 \rho$}
\RETURN $S$k$C(C_{1})$ and $S$k$C(C_{2})$
\ELSE
\RETURN $S$k$C(C_{1})$ and ($C_{2}$ as a single cluster)
\ENDIF
\ELSIF{$|\max_{i\in C_{2}}MCI_{i}-\min_{i\in C_{2}}MCI_{i}|>2 \rho$}
\RETURN $S$k$C(C_{2})$ and ($C_{1}$ as a single cluster)
\ELSE
\RETURN $C_{1}$ and $C_{2}$ as two clusters
\ENDIF
\end{algorithmic}
\end{algorithm}

This algorithm is commonly known as bisecting $k$-means clustering. Hence, S$k$C is a combination of load profile based clustering and bisecting clustering. While it obviously satisfies the local property to ensure smoothness, it may produce significantly more clusters than that needed.

\noindent \textbf{Remark}: In fact, we don't need to design such a complicated supervised $k$-means clustering. We should re-consider what should be the clustering criteria.

We adopt the notion of the end-to-end machine learning. The ultimate goal of clustering is for pricing. Hence, instead of conducting the $k$-means clustering based on load profiles, we can directly conduct the clustering based on each individual's $MCI$.

The shift in selecting the clustering criteria is significant. We can now directly guarantee the smoothness requirement with minimal room for market manipulation.

\subsection{Greedy Algorithm}

The idea is very simple. We first calculate the $MCI$ for all users and sort the tuple $(i, MCI_{i})$ in the ascending order of $MCI_{i}$. For notational simplicity, we assume $(i,MCI_{i})$ is already ordered, and the index $i$ simultaneously denotes user $i$ and its rank with respect to $MCI_{i}$. Then, starting from the first user, we use the interval of length $2\rho$ to cover the neighboring users and form the cluster. This procedure can be repeated until all users have been clustered. We provide the detailed algorithm in Algorithm $2$, and term this algorithm as greedy $k$-means clustering (or G$k$C in short).

\renewcommand{\algorithmicrequire}{\textbf{Input:}}  
\renewcommand{\algorithmicensure}{\textbf{Output:}} 

\begin{algorithm}[!ht]
\caption{Greedy $k$-means Clustering (G$k$C)}
\begin{algorithmic}[1]
\REQUIRE Ordered tuple $\left ( i,MCI_{i} \right ),i=1,...,n$
\ENSURE Clusters $C_{1},...,C_{\kappa}$
\STATE $i\leftarrow 1$
\STATE $k\leftarrow 1$
\WHILE{$i\neq n$}
\STATE $j=\arg\max_{m}MCI_{m},\,\,\,\,\,s.t. \  MCI_{m}\leq MCI_{i}+2\rho$
\STATE $C_{k}=\left \{ i,...,j \right \}$
\STATE $k\leftarrow k+1$
\STATE $i\leftarrow j+1$
\ENDWHILE
\STATE $\kappa=k$;
\RETURN $C_{1},...,C_{\kappa}$
\end{algorithmic}
\end{algorithm}

\subsection{Optimality of G$k$C}
This greedy algorithm achieves the minimal $\kappa$ to ensure the clustering criteria. Formally,

\vspace{0.1cm}
\noindent $\textbf{Theorem 2}$: The G$k$C selects the minimal number of clusters to ensure the clustering criteria:
\begin{equation}\label{ClusterCriteria}
    |MCI_{i}-p_{u(i)}|\leq \rho,\forall i.
\end{equation}

\noindent $\textbf{Proof}$:
Suppose the optimal clustering result is with fewer number of clusters $\kappa ^{\circ}$ (i.e., $\kappa ^{\circ}<\kappa$), which also ensures the clustering criteria (\ref{ClusterCriteria}).

Denote this clustering result by $\pi_{\circ}$ and the G$k$C clustering result by $\pi^{*}$. And denote for each cluster $k_{\circ} \in \pi_{\circ}$ ($k^{*} \in \pi^{*}$), the element with minimal $MCI$ by $s^{k_{\circ}}$ ($s^{k^{*}}$), and the element with maximal $MCI$ by $e^{k_{\circ}}$ ($e^{k^{*}}$).

Since $\pi_{\circ}$ is optimal, there don't exist clusters $i_{\circ}$, $j_{\circ}$, such that $s^{i_{\circ}}<s^{j_{\circ}}<e^{j_{\circ}}<e^{i_{\circ}}$ or $s^{i_{\circ}}<s^{j_{\circ}}<e^{i_{\circ}}<e^{j_{\circ}}$. That is, the clusters in $\pi_{\circ}$ won't overlap with each other. This can be straightforwardly proved by contradiction. We omit the proof due to space limit.

Hence, the clusters in $\pi_{\circ}$, just as the case for $\pi^{*}$, are also disjoint intervals.

Next, we examine the clusters in $\pi_{\circ}$ and $\pi^{*}$ along the axis of $MCI$. Since they are the clustering results for the same dataset, the first clusters in $\pi_{\circ}$ and $\pi^{*}$ share the same starting point, i.e., $s^{1_{\circ}}=s^{1^{*}}$.

Then, we observe the first difference between $\pi_{\circ}$ and $\pi^{*}$. We claim the first difference cannot be the starting point of some cluster. Otherwise, some profiles are left unclustered. 

Suppose the first difference happens at cluster $k_{\circ}$ and cluster $k^{*}$, then there are two possible conditions.

$(1)$ If $e^{k_{\circ}}>e^{k^{*}}$, then due to the construction process of G$k$C, $|s^{k_{\circ}}-e^{k_{\circ}}|>2\rho$, which won't satisfy the clustering requirement.

$(2)$ If $e^{k_{\circ}}<e^{k^{*}}$, then we can simply set $e^{k_{\circ}}=e^{k_{*}}$, $s^{k_{\circ}+1}=s^{k^{*}+1}$.

This forms a new clustering result with the same number of clusters in $\pi_{\circ}$. However, after the adjustment of $(2)$ along the axis of $MCI$, $\pi_{\circ}$ becomes exactly the same clustering result as $\pi^{*}$! This contradicts with our assumption that $\kappa^{\circ}<\kappa$. \hfill$\blacksquare$

\vspace{0.1cm}
\noindent \textbf{Remark}: The time complexity of G$k$C is $O(n\log n)$, which makes our data-driven pricing scheme even more practical. Since an algorithm with such a time complexity can be run even in real time, G$k$C solves the major concern of deploying our scheme: it requires the full knowledge of $p(t)$ to determine $MCI$. This implies that, the consumers are only aware of their electricity rates after the daily energy consumption. With the efficient G$k$C algorithm, based on advanced load forecasting methods, the microgrid operator can inform each user's estimated daily electricity rate in real time.

\section{Simulation Studies}\label{Simulaiton}

In this section, we first compare the performance of S$k$C and G$k$C to demonstrate the optimality of G$k$C. Then, we analyze how different parameters (e.g. $\rho$, $a$, etc) will affect the clustering results. This inspires us to further exploit more insights into the G$k$C results.

\begin{figure}[!t]
\centering
\includegraphics[width=3 in]{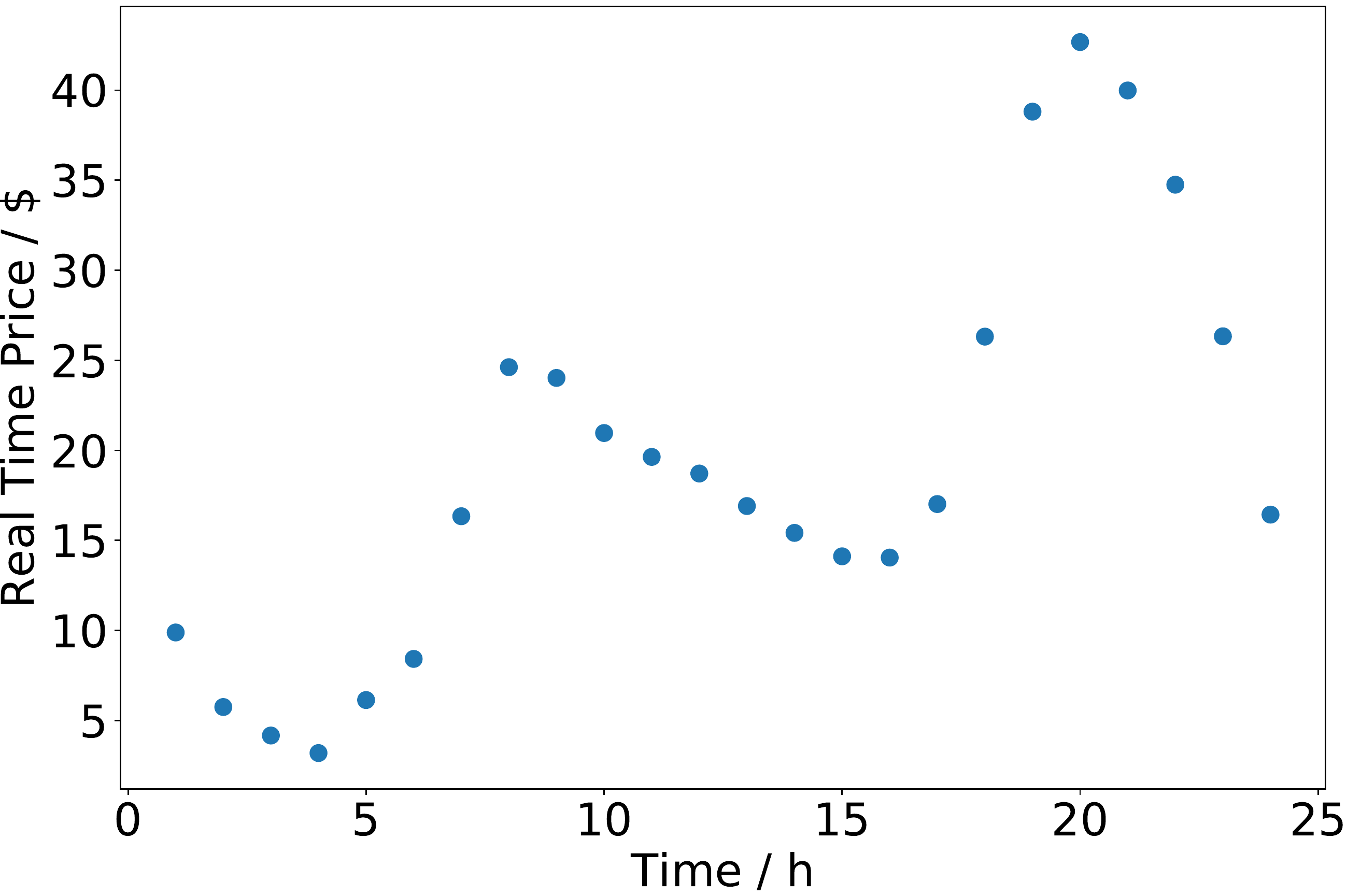}
\caption{Real Time Price Over a Day}
\label{RTP}
\end{figure}

Using the fitted price model, we select $a=0.00012$ and $b=-37.38$. According to the total load in system, we obtain the real time price of the day, which helps us to conduct the clustering. We visualize the real time price determined by Eq. (\ref{price}) in Fig. \ref{RTP}.

\subsection{S$k$C versus G$k$C}

\begin{figure}[!t]
\centering
\includegraphics[width=3 in]{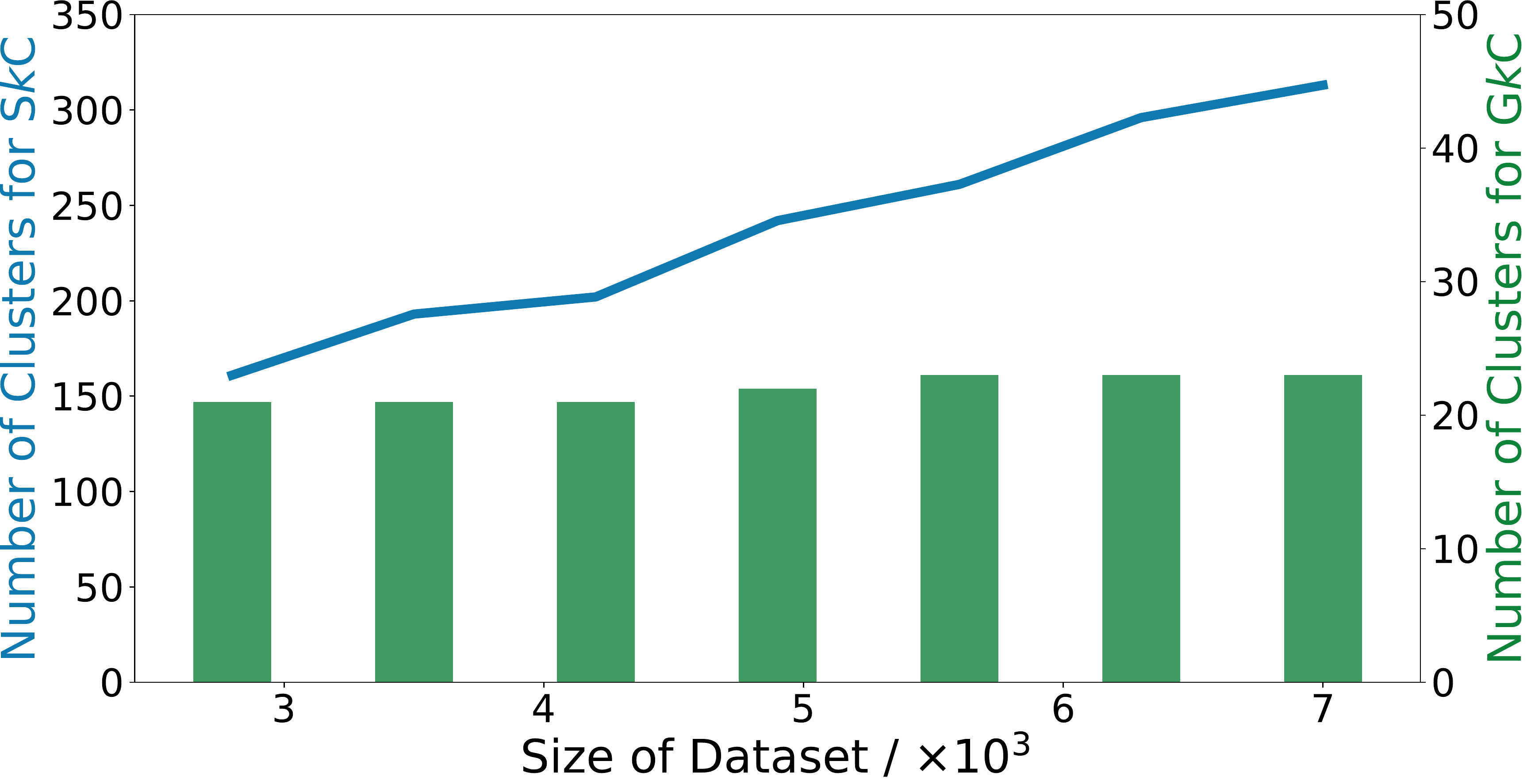}
\caption{Efficiency Comparison between S$k$C and G$k$C}
\label{DataK}
\end{figure}

\begin{figure}[!t]
\centering
\includegraphics[width=3 in]{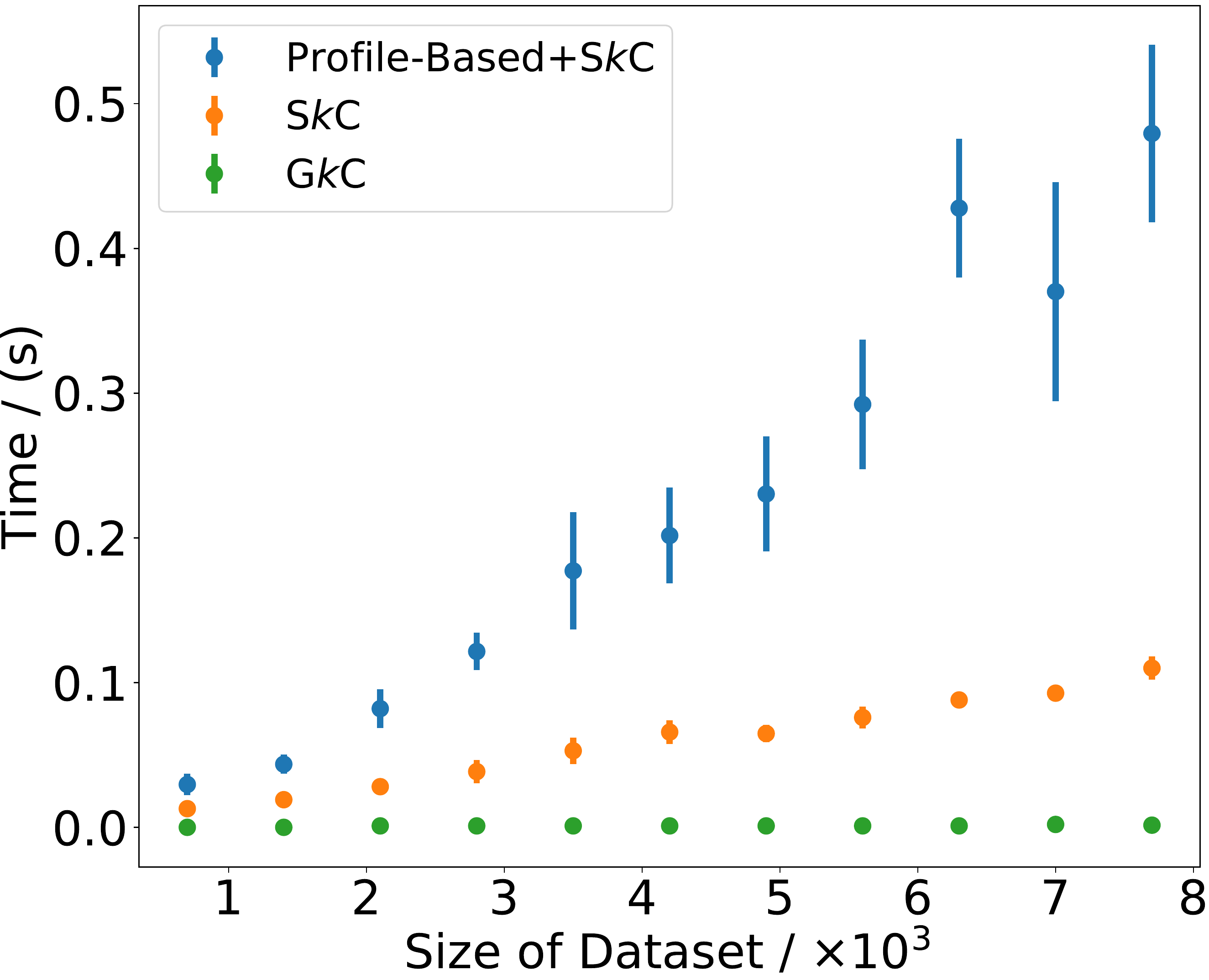}
\caption{Running Time Comparison between S$k$C and G$k$C (S$k$C only accounts for the time after obtaining the profile-based clustering results; profile-based+S$k$C measures the total time for the supervised $k$-means clustering).}
\label{DataTime}
\end{figure}

While both can guarantee the smoothness requirement, we compare the running time as well as the number of clusters returned by S$k$C and G$k$C. Obviously, due to the heuristic implementation of $k$-means clustering based on load profiles, G$k$C outperforms S$k$C in both aspects. 

In Fig. \ref{DataK}, with the increasing size of dataset, G$k$C produces a rather stable number of clustering, around $20$. In contrast, S$k$C produces hundreds of clusters and the number of clusters returned by S$k$C increases dramatically with the increasing size of dataset. Figure \ref{DataTime} compares the running time between S$k$C and G$k$C. Even if we don't take the time consumed by profile based clustering into account for S$k$C evaluation, G$k$C still runs faster than S$k$C.

In fact, as the dataset contains millions of profiles, G$k$C's running time is still $O(n\log n)$ yet S$k$C may suffer from the curse of dimensionality.

\begin{figure}[!t]
\centering
\includegraphics[width=3 in]{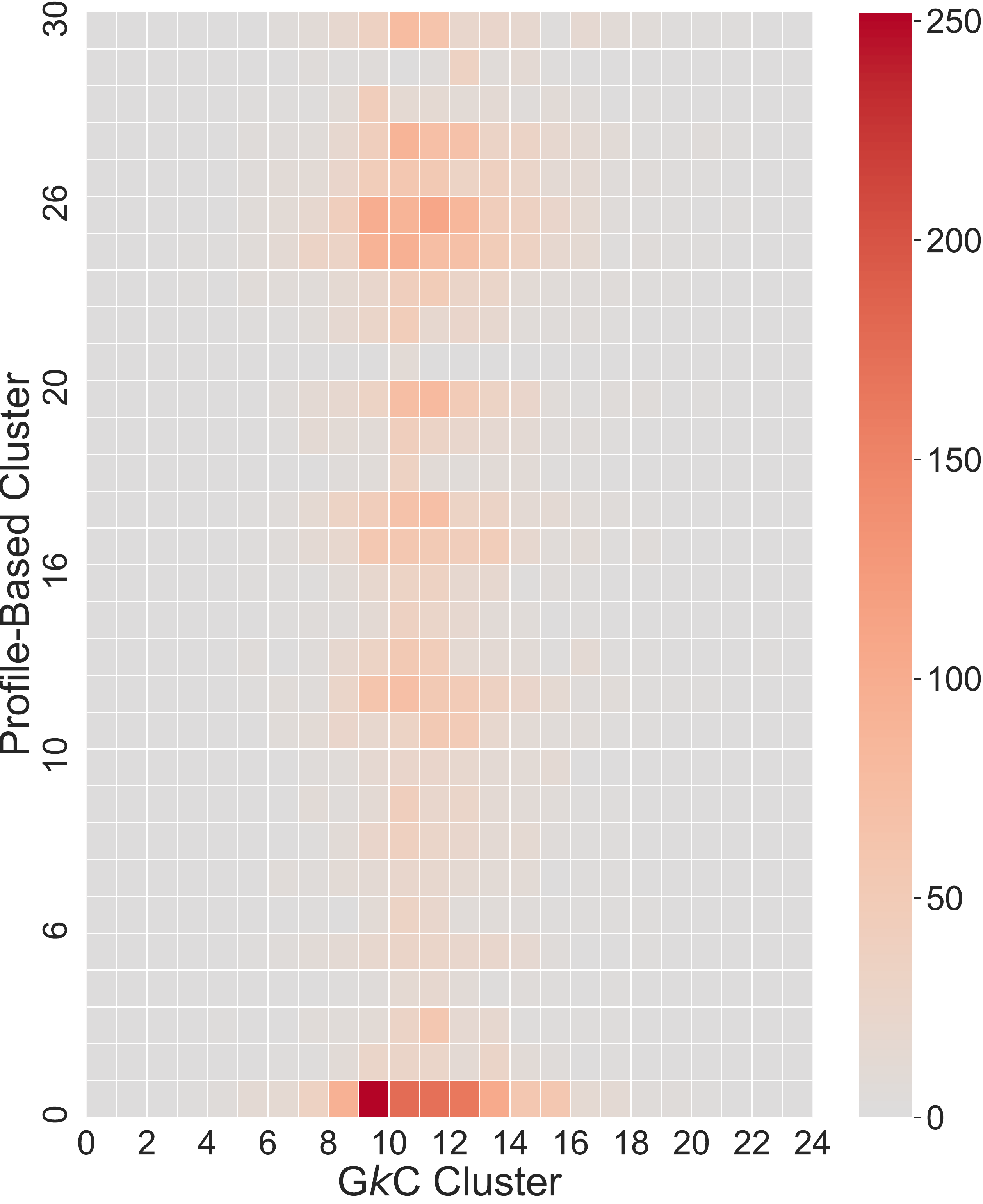}
\caption{Relationship Between G$k$C and Profile-Based Clustering when $\rho=0.5$ (the darker the rectangle, the more users in the rectangle).}
\label{G$k$CPBCHeatMap}
\end{figure}

To understand why S$k$C produces more clusters than G$k$C, we use Fig. \ref{G$k$CPBCHeatMap} for more insights. Since the initial clustering is conducted based on load profiles, most clusters need to be further decomposed into two or more clusters to ensure the clustering criteria (\ref{ClusterCriteria}). Hence, in Fig. \ref{G$k$CPBCHeatMap}, while the G$k$C only produces $24$ clusters, S$k$C needs to produce at least the same amount of clusters as the number of red colored rectangles.

\subsection{Dive into G$k$C}

We first examine the diversity of the load profiles in each G$k$C cluster. We define $\sigma_{k}$ to evaluate the mean distance to the cluster center for each cluster $k$:

\begin{equation}
    \sigma_{k}=\frac{\sum_{i\in C_{k}}||\mathbf{d}_{i}-\mathbf{c}_{k}||_{1}}{|C_{k}|},k=1,..K,
\end{equation}
where $\mathbf{c}_{k}$ denotes the center profile of cluster $k$ and $C_{k}$ denotes the set of users in cluster $k$, i.e., $C_{k}=\left \{ i|u(i)=k \right \}$.

\begin{figure}[!t]
\centering
\includegraphics[width=3 in]{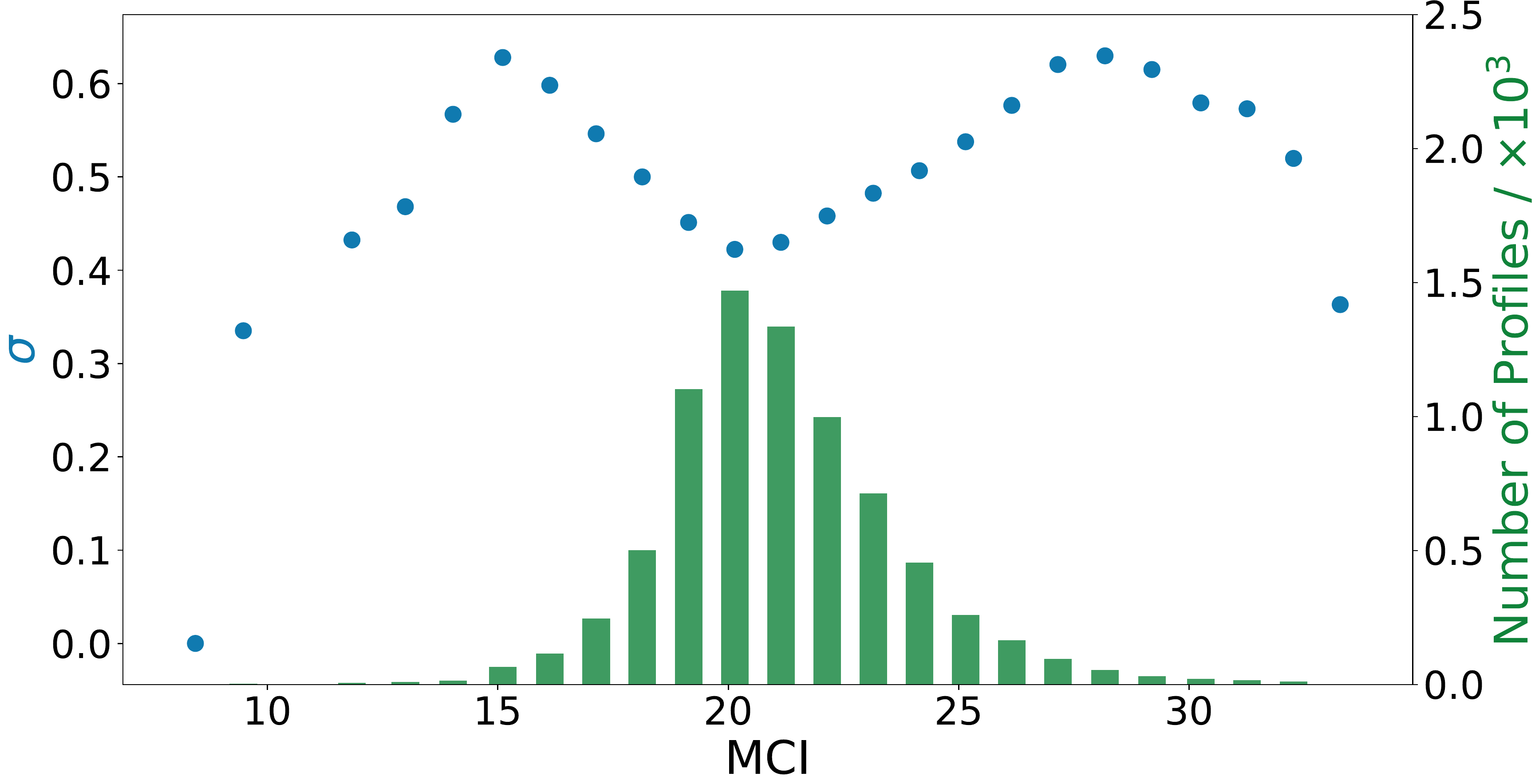}
\caption{Number of Clusters Based on Profiles and $\sigma$}
\label{MeanNum}
\end{figure}

We can observe in Fig. \ref{MeanNum} that the size of cluster seems like a binomial distribution: most users are around the medium $MCI$ in the dataset (the medium user is in cluster $12$). However, the diversity in each cluster has two peaks (cluster $6$ and cluster $19$). This is somewhat surprising. We conduct the profile-based $k$-means clustering for the two peaks for more insights.

\begin{figure}[!t]
\centering
\includegraphics[width=3 in]{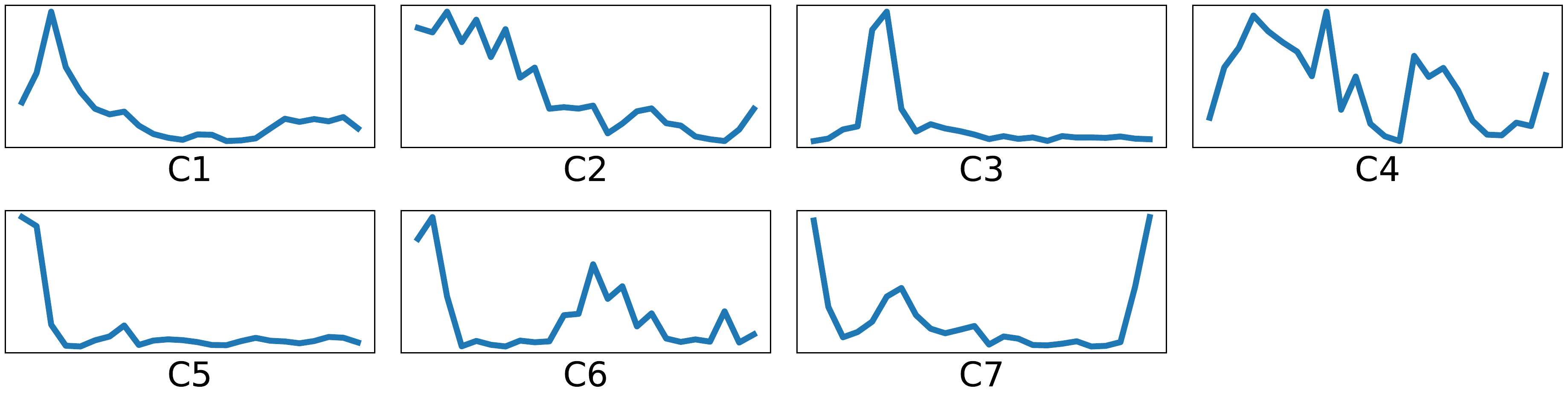}
\caption{Clustered Demand Profile of Cluster $6$ Based on $MCI$}
\label{C6Centers}
\end{figure}

\begin{figure}[!t]
\centering
\includegraphics[width=3 in]{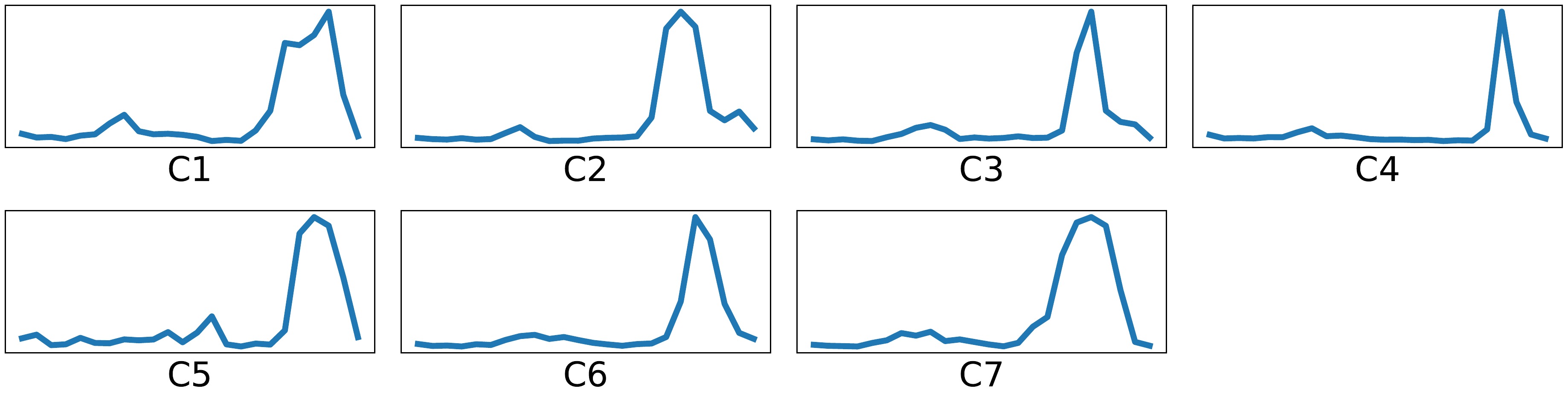}
\caption{Clustered Demand Profile of Cluster $19$ Based on $MCI$}
\label{C19Centers}
\end{figure}

Figure \ref{C6Centers} shows that the demand profiles in cluster $6$ are quite heterogeneous. In contrast, the demand profiles in cluster $19$ all have a peak in the evening (corresponding to the peak price in Fig. \ref{RTP}), as suggested by Fig. \ref{C19Centers}. This implies that for high $MCI$ users, since the high $MCI$ is caused by the peak price, their profiles tend to be similar. For these users, the results obtained by clustering based on load profiles and clustering based on $MCI$ are more likely to be aligned with each other. For other $MCI$ users, clustering based on $MCI$ is much more efficient!

\subsection{Sensitivity Analysis}
\begin{figure}[!t]
\centering
\includegraphics[width=3 in]{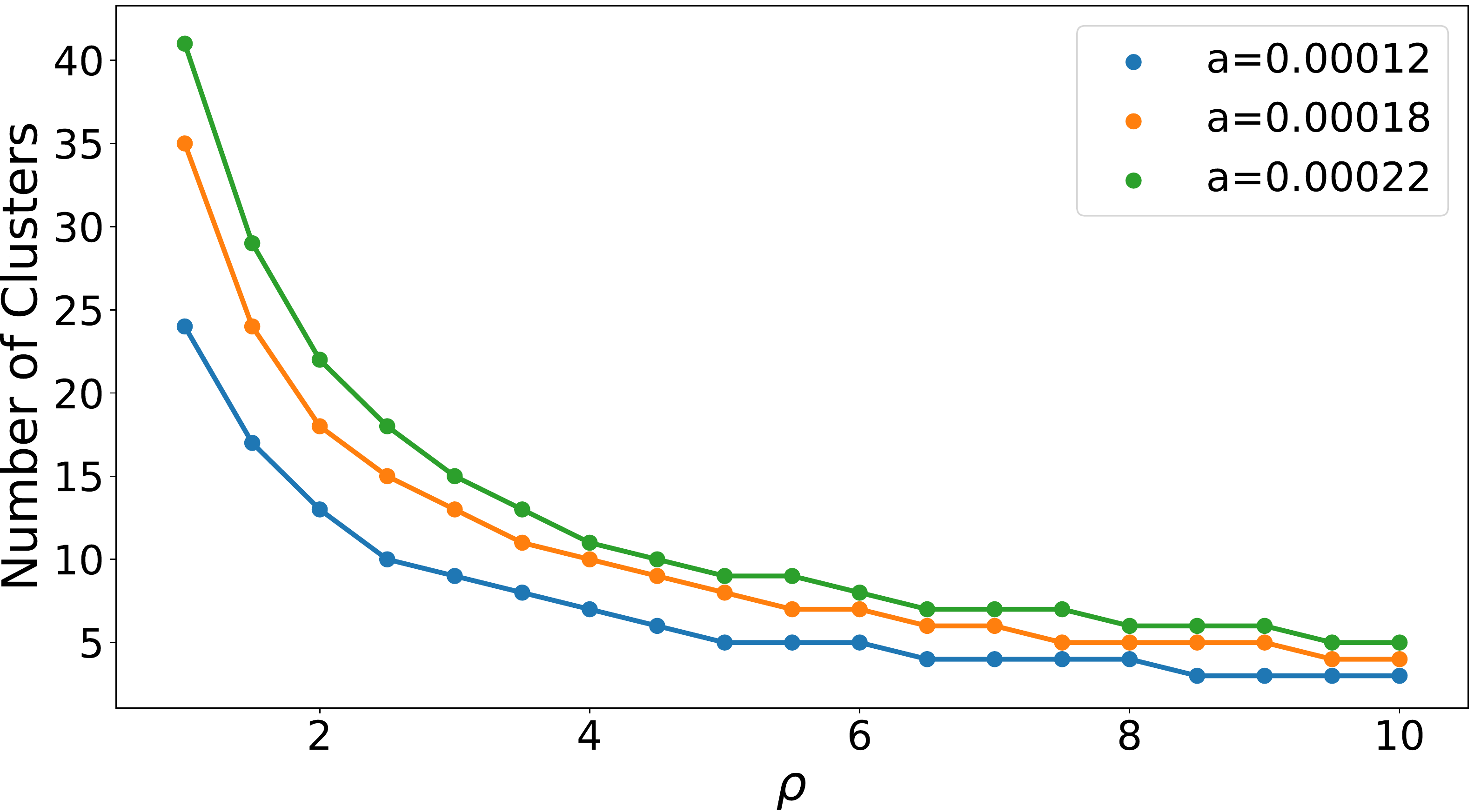}
\caption{Number of Clusters Evolving with $\rho$ for Different $a$}
\label{NumCiusterRho}
\end{figure}

We conclude this section by understanding the impacts of $\rho$ and $a$ on the clustering results. As suggested by Fig. \ref{NumCiusterRho}, the number of clusters produced by G$k$C diminishes quickly with the increasing of $\rho$. This also corresponds to our observation in Fig. \ref{MeanNum}, that the size of cluster distribution seems like a binomial distribution. Hence, the initial increase of $\rho$ leads to a more dramatic decrease in the number of clusters.

On the other hand, a larger $a$, which implies an even higher uniform electricity rate, leads to more clusters. Unlike the nonlinear impact of $\rho$, the impact of $a$ exhibits linearity: to twice the value of $a$ almost doubles the number of clusters.

\section{Conclusion}\label{Conclusion}
By adopting the notion of End-to-End machine learning, we identify the most efficient clustering criteria to design the data-driven profile-based pricing scheme. We further exploit the structure of the problem and the definition of smoothness, which inspire us to design an optimal and efficient greedy $k$-means clustering to implement the proposed pricing scheme.

This work can be extended in many ways. As we have mentioned in the remark in Section \ref{SystemModel}, it will be interesting to examine the possibility to relax the price taker assumption using a game theoretic analysis. We also intend to examine the temporal characteristics of user load profiles over a longer time span for a more detailed user $MCI$ profile characterization. This can enable the annual pricing plan design for the system operator.

\bibliographystyle{ieeetr}
\bibliography{reference}

\end{document}